\documentclass[superscriptaddress,showpacs,preprint]{revtex4-1} 
\usepackage{epsfig}
\usepackage{amsfonts}
\usepackage{amsmath}
\usepackage{xcolor}

\begin{document}

\title{Exact finite reduced density matrix and von 
Neumann entropy for the Calogero model}
\author{Omar Osenda}
\email{osenda@famaf.unc.edu.ar}
\affiliation{Facultad de Matem\'atica, Astronom\'{\i}a y F\'{\i}sica,
Universidad Nacional de C\'ordoba and IFEG-CONICET, Ciudad Universitaria,
X5016LAE C\'ordoba, Argentina}

\author{Federico M. Pont}
\email{pont@famaf.unc.edu.ar}
\affiliation{Facultad de Matem\'atica, Astronom\'{\i}a y F\'{\i}sica,
Universidad Nacional de C\'ordoba and IFEG-CONICET, Ciudad Universitaria,
X5016LAE C\'ordoba, Argentina}

\author{Anna Okopi\'nska }
\email{Anna.Okopinska@fuw.edu.pl}
\affiliation{Institute of Physics, Jan Kochanowski University, ul. 
\'Swi\c{e}tokrzyska 15, 25-406, Kielce, Poland}

\author{Pablo Serra}
\email{serra@famaf.unc.edu.ar}
\affiliation{Facultad de Matem\'atica, Astronom\'{\i}a y F\'{\i}sica,
Universidad Nacional de C\'ordoba and IFEG-CONICET, Ciudad Universitaria,
X5016LAE C\'ordoba, Argentina}

\begin{abstract}
The information content of continuous quantum variables systems is usually
studied using a number of well known approximation methods. The approximations
are made to obtain the spectrum, eigenfunctions  or the reduced density
matrices that are essential to calculate the entropy-like quantities that
quantify the information. Even in the sparse cases where the spectrum and
eigenfunctions are exactly known the entanglement spectrum, {\em i.e.} the
spectrum of the reduced density matrices that characterize the problem, must be
obtained in an approximate fashion. In this work, we obtain analytically a 
finite representation of the reduced density matrices of the fundamental state 
of the N-particle Calogero model for a discrete set of values of the 
interaction parameter. As a consequence, the exact entanglement spectrum and 
von Neumann entropy is worked out. 
\end{abstract}
\date{today}

\maketitle

\section{Introduction}

During the last few years, there has been an increasing interest in the
entanglement or, more generally, the information content of
quantum states of systems with continuous quantum variables (CV)
\cite{Chiribella2014,Braunstein2005,Cerf2007,Tichy2011}. This ample category
basically comprises, but is not exhausted by, electrons in atomic or molecular
systems and few particle model systems as the Moshinsky model
\cite{Moshinsky1968} and the Calogero model \cite{Calogero1969}. For an early
example of the application of information ideas to the internal continuous
variables of an artificial atom see the work of Amovilli and March
\cite{Amovilli2004}.

In this context, the amount of attention dedicated to the Moshinsky and, to a
lesser extent, the Calogero models seems at odds with their importance  or
the consequences that their study could have in understanding more realistic
systems. A casual onlooker would think so, but a more experienced one would
remember the scarcity of exact results about CV systems and think
otherwise. After all,  the study of entanglement in spin systems did
not reach its maturity until the exact formula to calculate the
entanglement of formation of an arbitrary state of two qubits was obtained by
Wootters \cite{Wootters1998}. This formula, together with the battery of
many-body exactly solvable models is the cornerstone of the amazing development
of the studies of entanglement in systems with finite Hilbert spaces
\cite{Amico2008}. In particular, the development of approximate methods has
benefited from the accumulation of benchmarks where they can be tested.

Incidentally, when the entanglement or information content of a quantum state is
under study, it is calculated using a plethora of entropy-like functions
\cite{Vedral2002,Nielsen2000}. Regrettably, the use of one or other is often
dictated by what it is feasible to calculate for a given system and not by the
requirements of some information task like it is mostly the case in spin
systems, where the entanglement of formation, the distillable entanglement
\cite{Bennett1996,Horodecki1998,Horodecki2001} and others come to mind.
In the same sense, the entanglement of formation of Gaussian states,
that are a genuine continuous variable system, can be determined
\cite{Giedke2003}. Anyway, the use of the von Neumann entropy and the
Jozsa-Robb-Wootters sub-entropy \cite{Jozsa1994} in CV systems is supported by
rigorous arguments \cite{Nichols2003} rather than numerical evidence from
particular systems. More recently, Iemini and Vianna 
\cite{Iemini2013}, have discussed how to compute the entanglement of 
indistinguishable particles pure states, both 
bosons and fermions, using the von Neumann entropy. On the other hand, in the 
work of Killoran {\em et al} \cite{Killoran2014} it is argued that any 
entanglement, even the one  that appears amongst identical particles due to 
symmetrization can be extracted and used for some task. All in all, the case 
for studying the von Neumann entropy for systems of identical particles is 
stronger than 
ever.

The technical difficulties associated to the calculation of informational
quantities in atomic-like systems have been discussed numerous times and
roughly speaking they can be classified in two kinds. The first kind is
related to the method employed to obtain the (approximate) quantum state and the
second kind is related to what informational quantities can be effectively
calculated from the quantum state, as has been said above. If the state is
obtained using a finite Hamiltonian approximation, like the ones that result
from the Hartree-Fock or the variational Ritz methods, there is no guarantee
that it will be manageable enough to obtain the reduced density matrices that
are necessary to calculate some entropy-like quantity, or that the approximate
quantum state even resembles the exact one or its entanglement content. From an
historical point of view it is interesting to note that this subject was the
motivation that led Moshinsky to study the model that is now called so after
him. Nevertheless, there has been a lot of progress in the study of CV
systems from more measures to quantify the information, as the geometric
entanglement \cite{Gottlieb2005} or the relative von Neumann entropy
\cite{Byczuk2012}, to  understand how the entanglement behaves when
approximate solutions obtained with the  Hartree-Fock method \cite{Zhang2014}
or the Ritz method \cite{Osenda2007,Ferron2009,Pont2010} are analyzed, and the
relationship between entanglement and energy in two-electron 
systems~\cite{Majtey2012}.

Despite that the Moshinsky and Calogero models have exact solutions, allowing to
obtain some of the required quantities to study the information content of
their quantum states, most studies about both models are restricted to the
so-called linear entropy, since it does not require a detailed knowledge of the
reduced density matrix spectrum, as is the case for the von Neumann entropy. If
the model considered has $N$ particles, then the $p$-particle reduced density matrix ($p$-RDM)
can be obtained by tracing out $N-p$  particles from the density matrix of
the whole system. 
Each of $p$-RDM can be used to obtain entropy-like quantities. The $p$-RDM has 
been obtained 
analytically for the N-particle Moshinsky system at arbitrary values of the 
interaction parameter~\cite{Pruski72}. In~\cite{ko13} the exact occupation 
numbers and the exact expression for the von Neumann p-entropies have been 
derived and their dependence on the interaction strength and the number of 
particles has been discussed.
 In this work we obtain
an exact finite analytical expression for the $p$-RDM for the $N$-particle 
one-dimensional Calogero model. We explicitly compute the cases for $p=1$ 
and $N$ up to $5$ for the totally symmetric (bosonic) ground state wave 
function. We analyze this case thoroughly, while the analysis of a totally 
anti-symmetric (fermionic) ground state function is restricted to the simplest 
case, {\em i.e.} two particles and their 1-RDM, since it proceeds similarly to 
the symmetric one. As we will
show,  in each case the entanglement spectrum \cite{Li2008,Calabrese2008} is
given by a finite number of eigenvalues if
the coefficient that characterizes the interaction between the particles assumes
certain particular values.  We analyze the relationship between our results with 
those found for the
Crandall and Hooke atoms  for $N=2$ \cite{Manzano2010} and for
 the N-particle Moshinsky model~\cite{ko13, Schilling, Benavides}.

\section{The model}\label{section:preliminary}

We adopt for the $N$-particle Calogero Hamiltonian \cite{Calogero1969} 
the expression of Sutherland \cite{s71}
\begin{equation}
\label{ehcal}
H = \sum_{i=1}^N\, h(i) +  \nu (\nu-1) \,\sum_{i< j}\,\frac{1}{(x_i-x_j)^2}\,,
\end{equation}
 where
\begin{equation}
\label{ehi}
h(i) = \frac{1}{2}\, p_i^2 + \frac{1}{2}\,\omega^2 x_i^2\, ,
\end{equation}
$p_i$ is the momentum operator and the masses are equal to one. In this reference 
 the author also gave the ground-state energy and the
corresponding  totally symmetric ground-state wave function,

\begin{equation}
\label{ewf0}
E\,=\,\left((N-1) \nu+1\right)\frac{N}{2}\,\omega  
\;\;\;;\;\;\;\Psi(x_1,\ldots,x_N)\,=\,C_{N,\nu}\,\Delta_{\nu}\,
\prod_{i=1}^N\,e^{-\frac{1}{2}\omega x_i^2}\,,
\end{equation}

\noindent where  $\Delta_{\nu}$
is the  Jastrow factor

\begin{equation}
\label{ejf}
\Delta_{\nu}\,=\, \prod_{i<j}\,\left|x_i-x_j\right|^{\nu}\,,
\end{equation}

\noindent and  $C_{N,\nu}$ is the normalization constant \cite{fw08},

\begin{equation}
\label{ecn}
C_{N,\nu}=\frac{2^{(N-1) N\nu/ 4}}{\pi^{N/4}}   \,\omega^{((N-1) \nu+1) N/4} 
\prod_{j=1}^{N}
\sqrt{\frac{\Gamma(1+ \nu)}{\Gamma(1+j \nu)}}  \,.
\end{equation}

Because the interaction potential of the Calogero model is a  homogeneous 
function of degree -2, as the kinetic energy, the rescaling 
$x\mapsto \sqrt{1\over {{\omega}}}x, \, E\mapsto { \omega E}$ maps the
problem to an
$\omega$-independent Schr\"odinger equations, in what follows we put $\omega=1$.

From the exact wave function for $N$ particles, $\Psi(x_1,\ldots,x_N)$, the
$p$-RDM is constructed as follows
\begin{eqnarray}\label{eprdm}
\rho^{(p)}_N(x_1,x_2,\ldots,x_{p};y_1,y_2,\ldots,y_{p} )&
=& \idotsint
 \; dx_{p+1}
dx_{p+2} \ldots dx_N \times \\ \nonumber
&&\Psi^{\star}(x_1,x_2,\ldots,x_{p},x_{p+1},\ldots,x_N)
\Psi(y_1,y_2,\ldots,y_p,x_{p+1},\ldots,x_N).
\end{eqnarray}

For $\nu=2 
n;\;n=1,\ldots$ the absolute value in equation~(\ref{ejf}) can be ignored and 
the only 
integrals needed to find $p$-RDM are Gaussian integrals with even powers in the 
Jastrow factor. Moreover, the $p$-RDM equation(~\ref{eprdm}) is then a 
multinomial 
expression of $(x_1,x_2,\ldots,x_{p};y_1,y_2,\ldots,y_{p} )$.
The general
expression for $\rho_N^{(p)}$ is quite cumbersome to obtain but it can be written
elegantly as a finite sum of Hermite functions.


\section{Expansion of the $p$-RDM in  Hermite Functions}

In order to obtain a general expression for the ground-state wave function
as an expansion on the orthonormal Hermite functions

\noindent 
\begin{equation}
\label{ehof}
\psi_k(x)\,=\,\frac{e^{-\frac{1}{2} x^2}\,H_k(x)}{\sqrt{2^k k! \pi^{1/2}}}\,,
\end{equation}

\noindent where $H_k(x)$ are the Hermite polynomials, we use the expression of
the Vandermonde Determinant \cite{lz10},

\begin{eqnarray}
\label{evandermondeH}
\prod_{i<j}^N \,(x_i-x_j) &=&
\left| \begin{array}{ccc}
x_1^0&\cdots&x_1^{N-1} \\
\vdots&\vdots& \vdots\\
x_N^0&\cdots&x_N^{N-1} \end{array} \right| \,=\,
\frac{1}{2^{(N-1) N/2}}\,\left| \begin{array}{ccc}
H_0(x_1)&\cdots&H_{N-1}(x_1)\\
\vdots&\vdots& \vdots\\
H_{0}(x_N)&\cdots&H_{N-1}(x_N) \end{array} \right|  \nonumber \\ \mbox{} &=&
\frac{1}{2^{(N-1) N/2}}\,
\sum_{i_1,\cdots ,i_N=0}^{N-1}
\varepsilon_{i_1,\cdots, i_N} H_{i_1}(x_1)  \cdots   H_{i_N}(x_N) \,,
\end{eqnarray}

\noindent where 
$\varepsilon_{i_1,\cdots ,i_N}$ is the completely antisymmetric tensor in the
indexes $0,\ldots,N-1$.

The ground-state wave function equations (\ref{ewf0}) and (\ref{ejf}) for $\nu=2 
n$
takes
the form

\begin{equation}
\label{ewf0H}
\begin{aligned}
\Psi_n^{(N)}(x_1,\ldots,x_N)\,=\,\frac{C_{N,2n}}{2^{n (N-1) N}}\,
e^{-\frac{1}{2} \sum_{i=0}^N x_i^2}\,\sum_{i_{1,1},\cdots,i_{1,N}=0}^{N-1}
\ldots
 \sum_{i_{2n,1},\cdots,i_{2n,N}=0}^{N-1} \varepsilon_{i_{1,1},\cdots , i_{1,N}}
\ldots
 \varepsilon_{i_{2n,1},\cdots , i_{2n,N}}\; \\
 \,H_{i_{1,1}}(x_1)\,\ldots\,H_{i_{2n,1}}(x_1) \,\ldots\
\,H_{i_{1,N}}(x_N)\,\ldots\,H_{i_{2n,N}}(x_N) \,.
\end{aligned}
\end{equation}

The product of Hermite polynomials can be written as a sum of Hermite
polynomials with known coefficients $A^{(q)}_{k_1,\cdots,k_m}$, which are given 
in  Ref.\cite{carlitz62}

\begin{equation}
\label{eprodH}
\prod_{m=1}^M\,H_{k_m}(x)\,=\,\sum_{q=0}^{\sum_m k_m} \,\frac{
A^{(q)}_{k_1,\cdots,k_m}}{\sqrt{2^{q} q! \pi^{1/2}}}
H_q(x)\,,
\end{equation}

\noindent where 

\begin{equation}
\label{eaq}
\begin{aligned}
A^{(q)}_{k_1,\cdots,k_m}\,=\, 
\sqrt{\frac{2^q \pi^{1/2}}{q!}}  \sum_{\stackrel{r_1,\cdots,r_m}{
r_1+\cdots+r_m\leq [\frac{q}{2}]}} \, (-1)^{\sum_i r_i} \frac{
\prod_i k_i!\; \left(\sum_i (k_i-2 r_i)\right)!}{\prod_i r_i! \prod_i (k_i-2 r_i)!
\left([\frac{q}{2}]-\sum_i r_i\right)!} \\
&\mbox{if} \;  q+k_1+\cdots+k_m \; \mbox{even}\,, \\
\mbox{}&\mbox{} \\
\end{aligned} 
 \,,
\end{equation}

\noindent and $A^{(q)}_{k_1,\cdots,k_m}\,=\,0$ if $q+k_1+\cdots+k_m $ odd.
Then, inserting  equations (\ref{ehof}) and (\ref{eprodH}) into  equation  
(\ref{ewf0H})
we obtain

\begin{equation}
\label{ewfH}
\begin{aligned}
\Psi_n^{(N)}(x_1,\ldots,x_N)\,=\,\frac{C_{N,2n}}{2^{n (N-1) N}}\,
\sum_{i_{1,1},\cdots,i_{1,N}=0}^{N-1} \ldots
 \sum_{i_{2n,1},\cdots,i_{2n,N}=0}^{N-1} \varepsilon_{i_{1,1},\cdots , i_{1,N}}
\ldots
 \varepsilon_{i_{2n,1},\cdots , i_{2n,N}}\; \\
\prod_{k=1}^{N}\,\sum_{q_k=0}^{\sum_{l=1}^{2 n} i_{l,k}}
A^{(q_k)}_{i_{1,k},\cdots,i_{2 n,k}}
\,\psi_{q_k}(x_k)\,,
\end{aligned}
\end{equation}

\noindent equation (\ref{ewfH}) is
the
expansion of the ground-state wave function in the orthonormal Hermite basis, 
which, in a compact way can be written as

\begin{equation}
\label{ewfHc}
\Psi_n^{(N)}(x_1,\ldots,x_N)\,=\,\sum_{q_1,\cdots ,q_N=0}^{2 n
(N-1)}\;a_{q_1,\cdots ,q_N}\;
\psi_{{q_1}}(x_1) \ldots \psi_{{q_N}}(x_N) \,,
\end{equation}

\noindent where, from equation (\ref{ewfH}),

\begin{equation}
\label{ean}
\begin{aligned}
a_{q_1,\cdots ,q_N}\,=\,\frac{C_{N,2n}}{2^{n (N-1) N}}\,
\sum_{\stackrel{
i_{1,1},\cdots,i_{1,N}=0}{ \sum i_{n_i,1}\geq q_1,\cdots}}^{N-1} 
\ldots
 \sum_{\stackrel{i_{2n,1},\cdots,i_{2n,N}=0}{\cdots ,\sum i_{n_i,N}\geq
q_N}}^{N-1} 
\varepsilon_{i_{1,1},\cdots , i_{1,N}} \ldots
 \varepsilon_{i_{2n,1},\cdots , i_{2n,N}}\; \\
\prod_{k=1}^{N}\, A^{(q_k)}_{i_{1,k},\cdots,i_{2n,k}} \,.
\,
\end{aligned}
\end{equation}

From equations (\ref{eprdm}), (\ref{ewfHc}) and (\ref{ean}), the expression
of the $p-$RDM takes the simple form

\begin{eqnarray}
\label{enrdm}
\rho_N^{(p)}(x_1,\ldots,x_{p}; y_1,\ldots,y_{p})& = &\sum_{q_1,\cdots ,q_N=0}^{2 n
(N-1)}\;
\sum_{r_1,\cdots ,r_N=0}^{2 n (N-1)}\; a_{q_1,\cdots ,q_N}\; a_{r_1,\cdots
,r_N}\;
\delta_{q_{p+1},r_{p+1}}\cdots \delta_{q_{N},r_{N}}  \nonumber\\
\mbox{} & \mbox{} &
\psi_{q_1}(x_1) \cdots \psi_{q_{p}}(x_{p}) \psi_{r_1}(y_1) \cdots 
\psi_{r_{p}}(y_{p}) \nonumber \\
\mbox{} & \mbox{} &  \\
\mbox{}& = & 
\sum_{q_1,\cdots ,q_N=0}^{2 n (N-1)}\;\sum_{r_1,\cdots ,r_{p}=0}^{2 n (N-1)}\;
a_{q_1,\cdots ,q_N}\; a_{r_1,\cdots ,r_{p},q_{p+1} ,q_N} \nonumber \\
\mbox{}& \mbox{} &
\psi_{q_1}(x_1) \cdots \psi_{q_{p}}(x_{p}) \psi_{r_1}(y_1) \cdots
\psi_{r_{p}}(y_{p}) \,. \nonumber
\end{eqnarray}

\noindent With this expression the $p$-RDM is a symmetric kernel given by a 
finite sum of terms which are the product of the orthonormal functions 
$ \psi_{q_1}(x_1) \cdots \psi_{q_{p}}(x_{p})$. Therefore \cite{mikhlin}, these 
products are also the eigenfunctions of 
$\rho_N^{(p)}$, and  
its spectrum  is given by the eigenvalues of the 
$\left(2 n (N-1)+1\right)^{p} \times \left(2 n (N-1)+1\right)^{p}$ matrix

\begin{equation}
\label{espectro}
[\rho_N^{(p)}]_{i,j}\,=\sum_{q_{p+1},\cdots ,q_N=0}^{2 n (N-1)}\;a_{q_1,\cdots
,q_N}\; a_{r_1,\cdots ,r_{p},q_{p+1} ,q_N} \,,
\end{equation}

\noindent where the indexes $i,j$ are related to the indexes $q,r$ by 

\begin{eqnarray}
\label{eindex}
i=1+q_1+\left(2 n (N-1)+1\right) q_2+\cdots+\left(2 n (N-1)+1\right)^{p-1} q_p
\nonumber \\
j=1+r_1+\left(2 n (N-1)+1\right) r_2+\cdots+\left(2 n (N-1)+1\right)^{p-1}
r_p\,.
\end{eqnarray}
Because the Hermite functions have a definite parity, the matrix elements of the 
$p$-RDM  equation(\ref{espectro}) are zero for odd values of $i+j$.

Once the eigenvalues $\lambda_k$ of the matrix (\ref{espectro}) are calculated, the
von Neumann and
linear  entropies of the p-RDM of the N-particle Calogero model 
with the interaction strength $\nu(\nu-1)=2n(2n-1)$ can be obtained as

\begin{equation}
\label{ees}
S_N\,=\,-\sum_{k=1}^{(2 n (N-1)+1)^p}\,\lambda_k\,\log_2(\lambda_k) \;\;;\;\;
L_N\,=\,1-\sum_{k=1}^{(2 n (N-1)+1)^p}\,\lambda_k^2 \,.
\end{equation}


\section{The two-boson Calogero model}

For $N=2$ and $\nu=2 n$ the ground-state wave function equation(\ref{ewf0}) 
takes the form

\begin{equation}
\label{ewf02n}
\Psi_n^{(2)}(x_1,x_2)\,=\,\frac{2^n}{\sqrt{\pi}} \sqrt{(2 n)!\over (4 n)!}\,
(x_1-x_2)^{2 n}\,e^{-\frac{1}{2}
(x_1^2+x_2^2)}\,.
\end{equation}

By virtue  of  equation(\ref{enrdm}), the $1$-RDM can be written as a finite
expansion in
a
bi-orthogonal basis set, 

\begin{equation}
\label{erho2d}
\rho^{(1)}_2({x};{y})\,=\, \sum_{i,j=0}^{2
n}\,\rho_{i,j}\,\psi_i({x})\,\psi_j({y})\,.
\end{equation}

Once $\rho^{(1)}_2$ has been  obtained we have to solve the eigenvalue problem
\begin{equation}
\label{eevr}
\int_{-\infty}^\infty\,
\rho^{(1)}_2(x_1,x_2)\,\varphi_i(x_2)\,dx_2\,=\,\lambda_i\,\varphi_i(x_1)\,.
\end{equation}

\begin{figure}
\begin{center}
\includegraphics[width=6.cm]{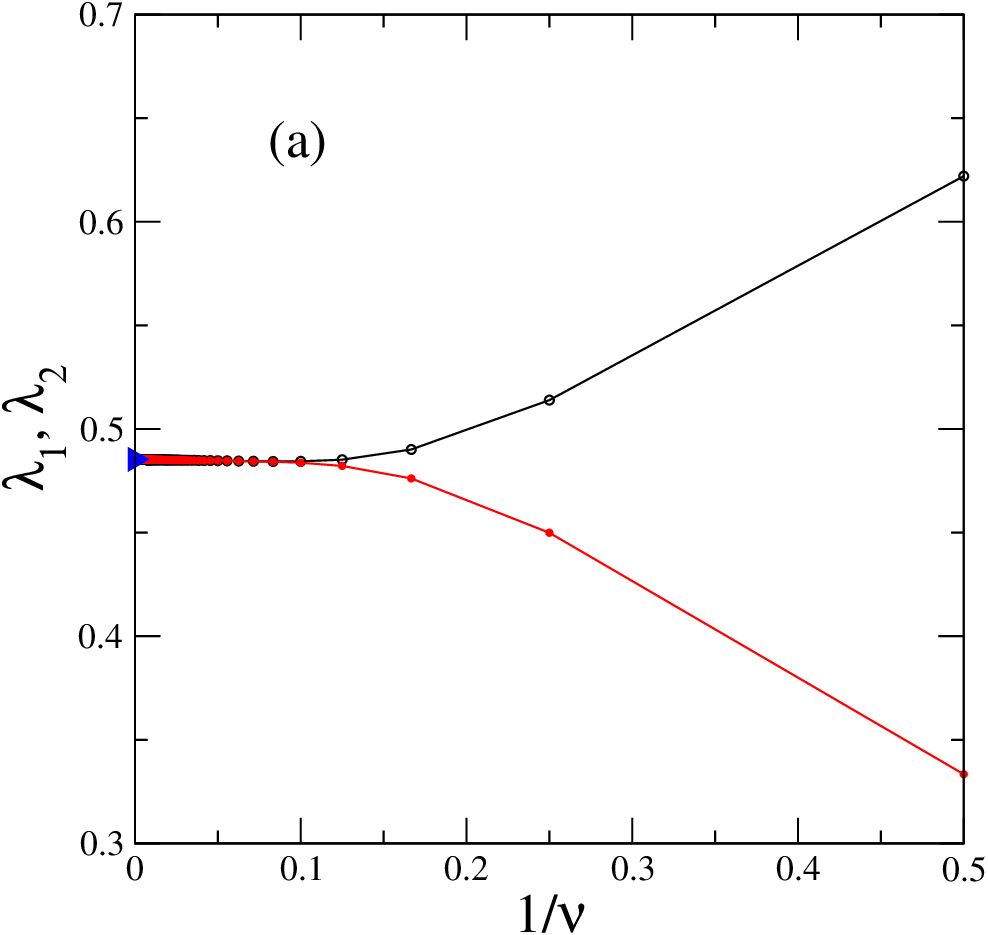} \hspace{.2cm}
\includegraphics[width=6.cm]{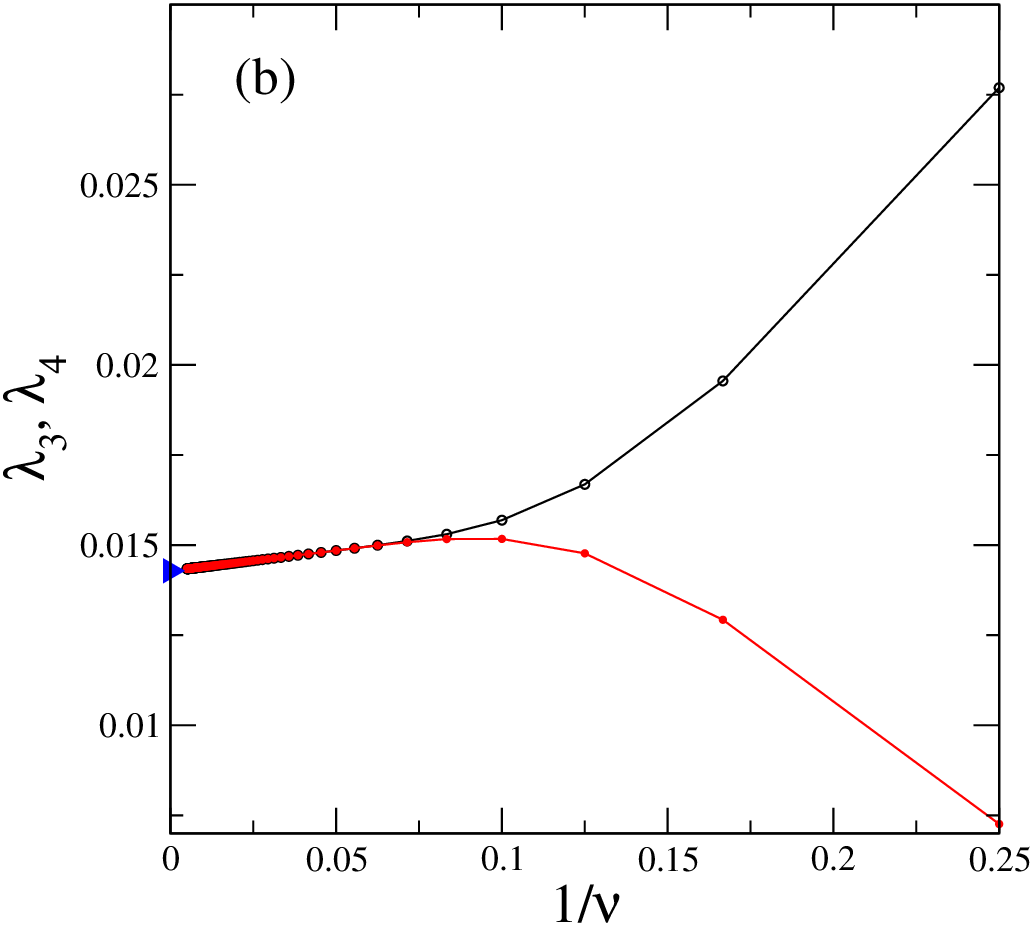}
\end{center}
\caption{\label{favN2} a) First (black line) and second (red line) eigenvalues of  
$\rho_2^{(1)}$ for 
$N=2$ 
as a function of $1/\nu$.  b) Third (black line)  and fourth (red line) eigenvalues of  $\rho_2^{(1)}$
for $N=2$
as a function of $1/\nu$.  The asymptotic degenerated values are shown by a blue triangle}
\end{figure}

Note that the $(2 n+1)\times (2 n +1)$ 1-RDM is a real symmetric matrix 
with two blocks, one $(n+1)\times (n+1)$ even
block 
and one $n \times n$ odd block.
For the case $n=1 \Rightarrow \nu=2$ the matrices are  $2 \times 2$  and
$1 \times 1$ respectively, and its entries can be calculated analytically
using equations (\ref{eprodH}-\ref{eindex}). The normalization constant is 
$C_{2,2}\,=\,\frac{1}{\sqrt{3 \pi}}\,$,
the non-zero 
coefficients for the Hermite expansion equation (\ref{eprodH}) are

\begin{equation}
\label{enzc}
A^{(0)}_{0,0}\,=\,\pi^{1/4} \;;\; A^{(0)}_{1,1}\,=\,2 \pi^{1/4} \;;\;
A^{(1)}_{1,0}\,=\,A^{(1)}_{0,1}\,=\,\sqrt{2} \pi^{1/4} \;;\; 
A^{(2)}_{1,1}\,=\,\sqrt{8}  \pi^{1/4} \,,
\end{equation}

\noindent all the others are zero. 
Then, from equation (\ref{ean}), the non-zero coefficients of the
wave-function 
expansion equation(\ref{ewfHc}) are

\begin{equation}
\label{ea21}
a_{0,0}\,=\,\frac{1}{\sqrt{3}},\;a_{0,2}\,=\,a_{2,0}\,=\,\frac{1}{\sqrt{6}},\;
a_{1,1}\,=\,\frac{1}{\sqrt{3}} \,.
\end{equation}
The complete matrix is

\begin{equation}
\label{ern2}
\left[\rho_2^{(1)}\right]\,=\,\left( \begin{array}{ccc}
 \frac{1}{2}&0&\frac{1}{3 \sqrt{2}}\\
 0&\frac{1}{3}&0\\
 \frac{1}{3 \sqrt{2}}&0&\frac{1}{6}\end{array} \right) \,,
\end{equation}

whose eigenvalues are

\begin{equation}
\label{eavn2}
\lambda_\pm\,=\,\frac{2\pm\sqrt{3}}{6}\;;\;\lambda_2\,=\,\frac{1}{3}\,.
\end{equation}

Of course, for $n>1$, even though the $1$-RDM was obtained analytically, the
eigenvalues must be calculated numerically.

In figure \ref{favN2}a) the two largest eigenvalues of $\rho$
are shown against $1/\nu$, and in figure  \ref{favN2}b) the third and fourth
eigenvalues.

In the strong interaction limit $\nu\to \infty$ the eigenvalues become doubly degenerate and can be calculated within the harmonic approximation~\cite{HA2015} (which becomes exact in this limit) to be given by asymptotic formulas
\begin{equation}
\lambda_{2k+1,2k+2}^{\nu\rightarrow\infty}=2\sqrt{2}(3-2\sqrt{2})(17-12\sqrt{2})^k,~~~~k=0,1,\ldots.
 \label{occ}
\end{equation}
For the lowest occupancies we obtain $\lambda_{1}^{\nu\rightarrow\infty}=\lambda_{2}^{\nu \rightarrow\infty}=2\sqrt{2}(3-2\sqrt{2}) \approx 0.485281$, $\lambda_{3}^{\nu\rightarrow\infty}=\lambda_{4}^{\nu \rightarrow\infty}=2\sqrt{2}(99-70)\sqrt{2}) \approx 0.0142853$. 
The eigenvalues $\lambda_1,\;\lambda_2$ coincide to  15  digits
((real(8) precision) for $n\ge22$, and  $\lambda_3,\;\lambda_4$ for $n\ge25$ with the asymptotic values. 

\begin{figure}
\begin{center}
\includegraphics[width=7.cm]{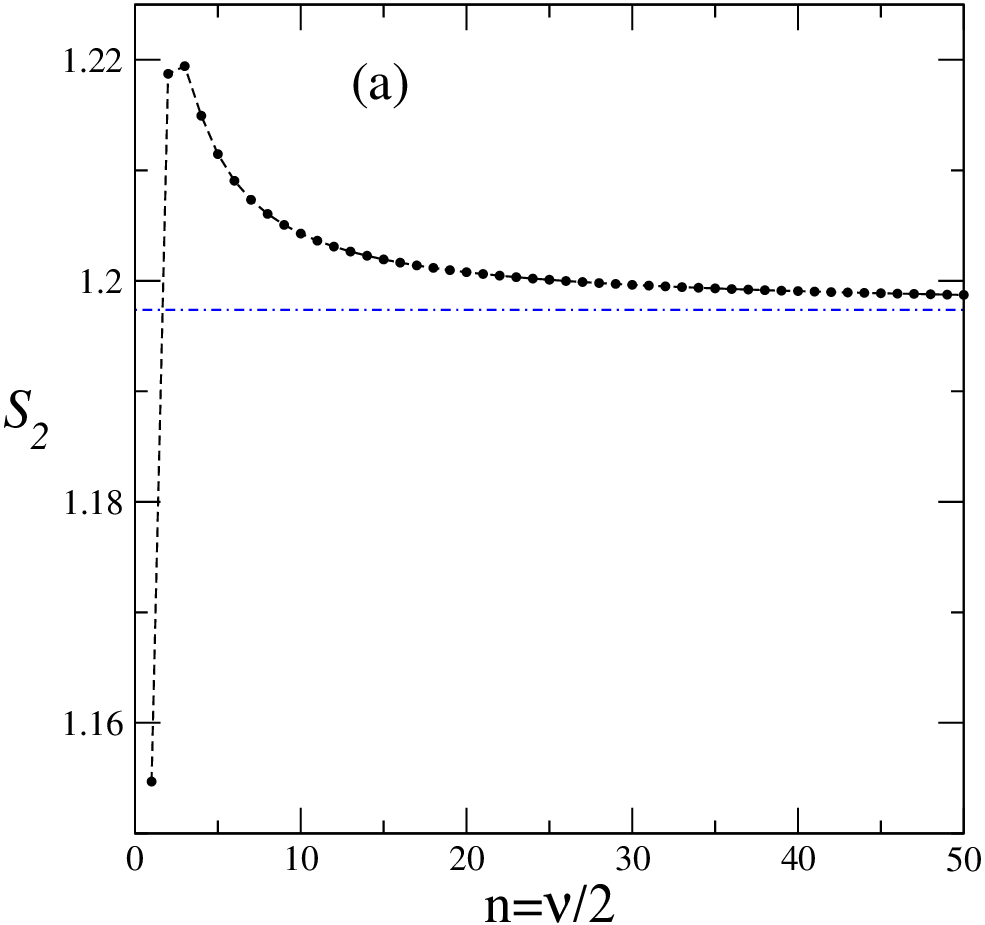}
\includegraphics[width=7.cm]{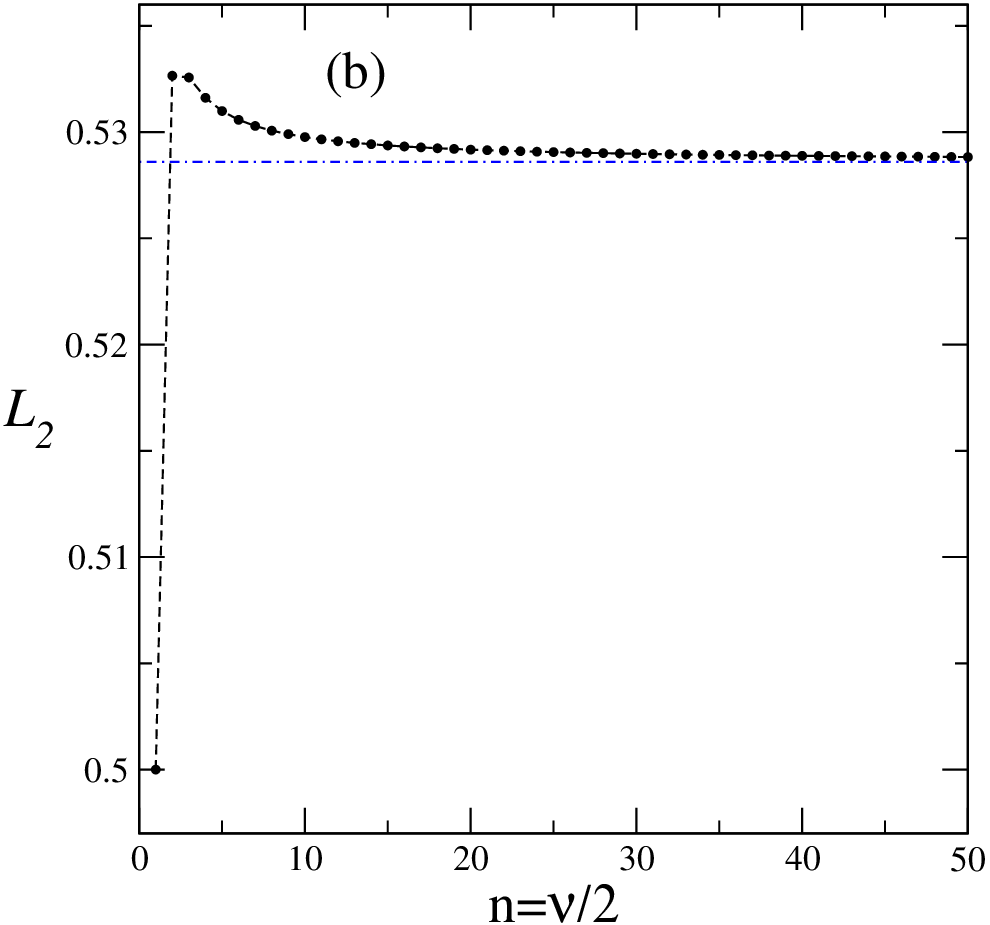}
\end{center}
\caption{  \label{fvne} a) von Neumann entropy for $N=2$. b) Linear  entropy for
$N=2$. The asymptotic values are indicated by the dash-dotted blue lines.}
\end{figure}

The  von Neumann entropy is shown in figure \ref{fvne} a), and the linear
entropy in
figure \ref{fvne}b) for $n=1,\ldots,50$. We note that the entropies have a
maximum between 
$n=2$ and $n=3$. 
Using the analytical formula for
$\lambda_{k}^{\nu\rightarrow\infty}$~(\ref{occ}) and performing the summation in the formulas~(\ref{ees}), the asymptotic values of the entropies can be calculated. 
The asymptotic value of the linear entropy is determined as

\begin{equation}
L_2^{^{\nu\rightarrow\infty}}=1-2\sum_{k=1}^{\infty}[\lambda_{k}^{\nu\rightarrow\infty}]^2= 1-{\sqrt{2}\over 3} \approx 0.528595,
\end{equation}

\noindent and the asymptotic value of the von Neumann entropy as being equal to

\begin{equation} S_{2}^{\nu\rightarrow\infty}=-2\sum_{k=1}^{\infty}\lambda_{k}^{^{\nu\rightarrow\infty}}
\mbox{log}_{2}\lambda_{k}^{^{\nu\rightarrow\infty}}=\frac{3 \log_{2} \left(3+2 \sqrt{2}\right)}{2 \sqrt{2}}-\frac{3}{2} \approx 1.19737.
\end{equation}

\noindent The exact values plotted in figure \ref{fvne} approach nicely those limits.

\begin{figure}
\begin{center}
\includegraphics[width=7.cm]{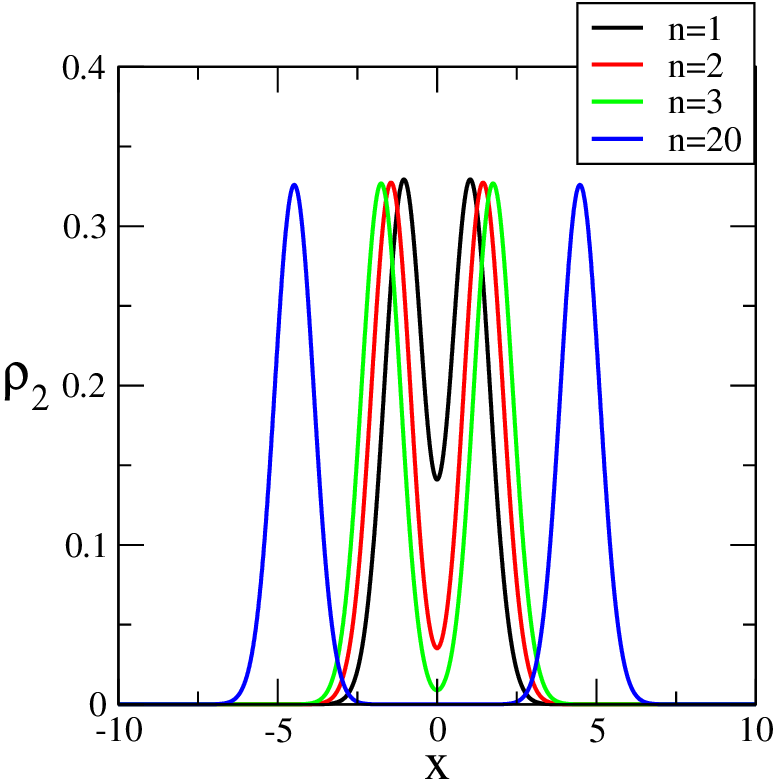}
\end{center}
\caption{  \label{frho2} One-particle density $\rho_2(x)$ for  $N=2$ and $n=1, 
2, 3, 20$.}
\end{figure}

Finally, in figure \ref{frho2} we show the one-particle density for 
$n=1, 2, 3, 20$. The one-particle density  is an even function of the 
position, 
as can be expected from the symmetries of the Hamiltonian.
The two-peaked one-particle density explains, at some extent, the behavior of
the largest eigenvalues of the entanglement spectrum, as shown in
figure~\ref{favN2}, since for large enough values of $n$ it is easy to envisage
that the eigenfunctions of the reduced density matrix that corresponds to the
two largest eigenvalues should be, approximately,  two peaked functions whose
peaks should coincide with the peaks of the RDM, one of the then 
even and the
other one odd and that for large enough $n$ both functions should weight more or
less the same when the spectral decomposition of the RDM is considered. This 
reasoning applies when the two peaks of the one-particle density are well 
separated 
by a region where its magnitude is negligible. Of course, this is not the case 
for small values of $\nu$, where the height of the two peaks is not so 
different from the value of the particle density at the origin. Note that, 
because the wave function vanishes for $x_{1}=x_{2}$, the one-particle density 
has two peaks even in the limit $\nu\rightarrow 1$.

\section{The three, four and five-boson Calogero model}

\begin{figure}
\begin{center}
\includegraphics[width=7.cm]{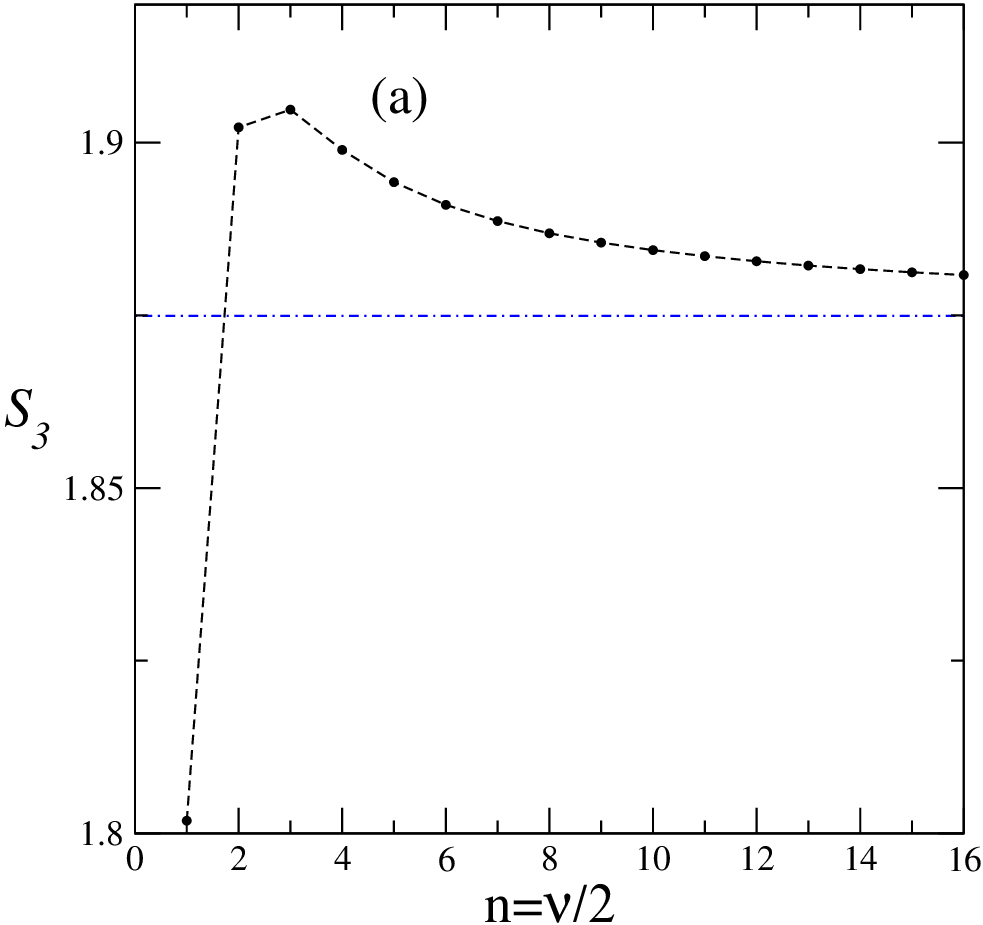}
\includegraphics[width=7.cm]{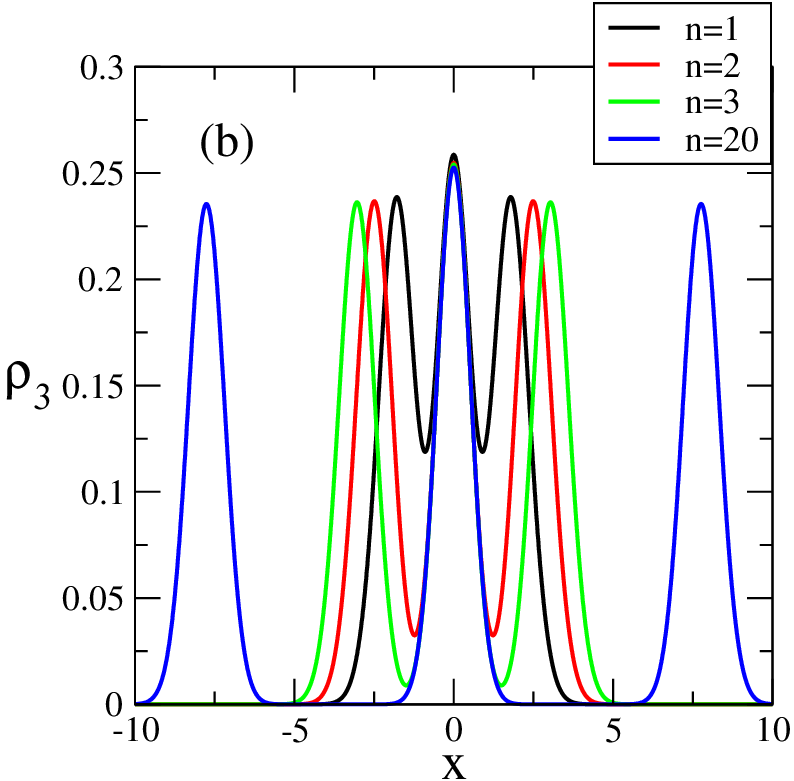}
\caption{\label{fvneN3} a) von Neumann entropy for $N=3$, the asymptotic value 
is indicated by the dash-dotted blue line. 
 b) One-particle
density $\rho_3(x)$ for  $N=3$ and $n=1, 2, 3, 20$ }
\end{center}
\end{figure}

For $N>2$, even though the $p$-RDM was obtained analytically, the
eigenvalues must
be calculated numerically. Despite its elegant and concise form, the evaluation of all the terms involved 
in equation \ref{espectro} for increasing values of $N$ becomes very demanding, 
so 
we present results for $N$ up to five.

Figure \ref{fvneN3}a)  
shows the von Neumann entropy for N=3 and its asymptotic value 
\mbox{$S_{3}^{\nu\rightarrow\infty} \approx 1.87494$} \cite{HA2015,HApc}, and 
figure \ref{fvneN3}b) the
one-particle density for $n=1,2,3, 20$.

As can be easily appreciated from figure~\ref{fvneN3}a), the behavior of the
von
Neumann entropy is quite similar to the behavior already found for the case
with $N=2$. Once again the most easily recognizable  feature is the maximum
that is attained near $n=3$. Figure~\ref{fvneN3}b) shows the particle density
as a function of $n$. As can be expected, the particle density has three well
defined peaks, one of them centered in the origin of the coordinate axis and
the other two placed symmetrically to both sides of the origin. The particle
density is broader, as a function of the coordinate, for three particles than
for only two, reflecting the fact that the repulsive term is stronger because
the extra particle.

\begin{figure}
\begin{center}
\includegraphics[width=7.cm]{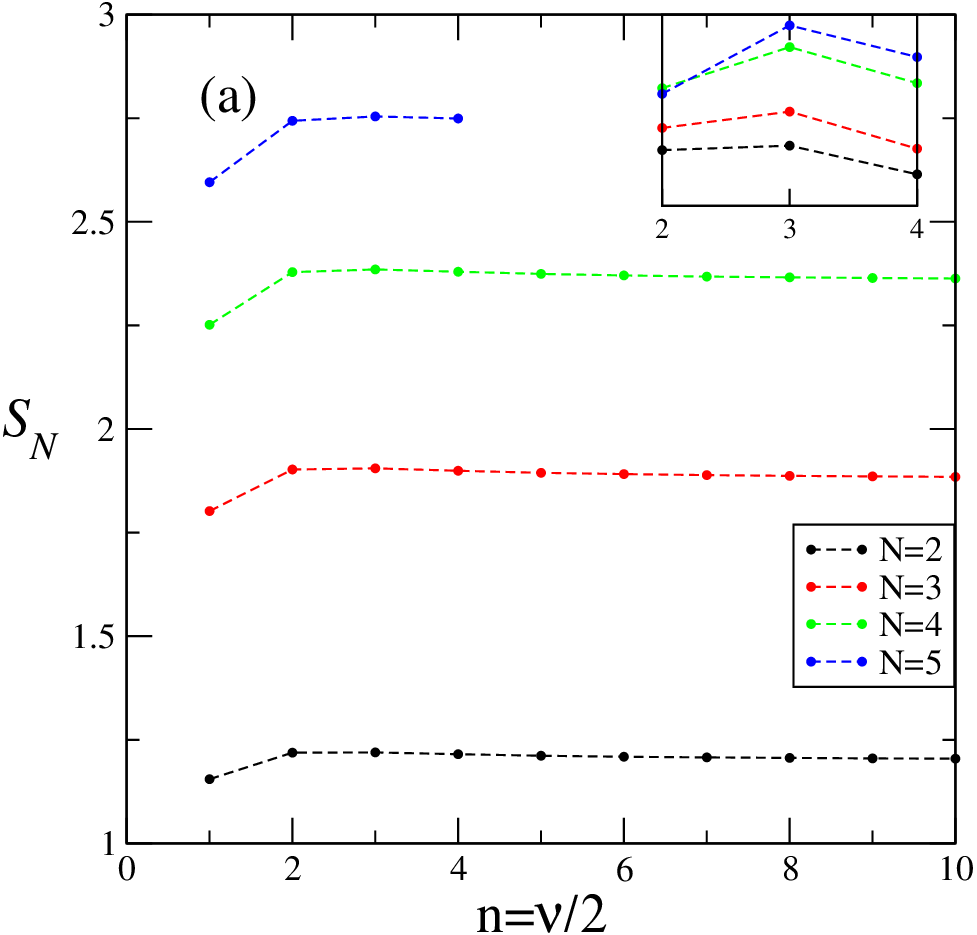}
\includegraphics[width=7.cm]{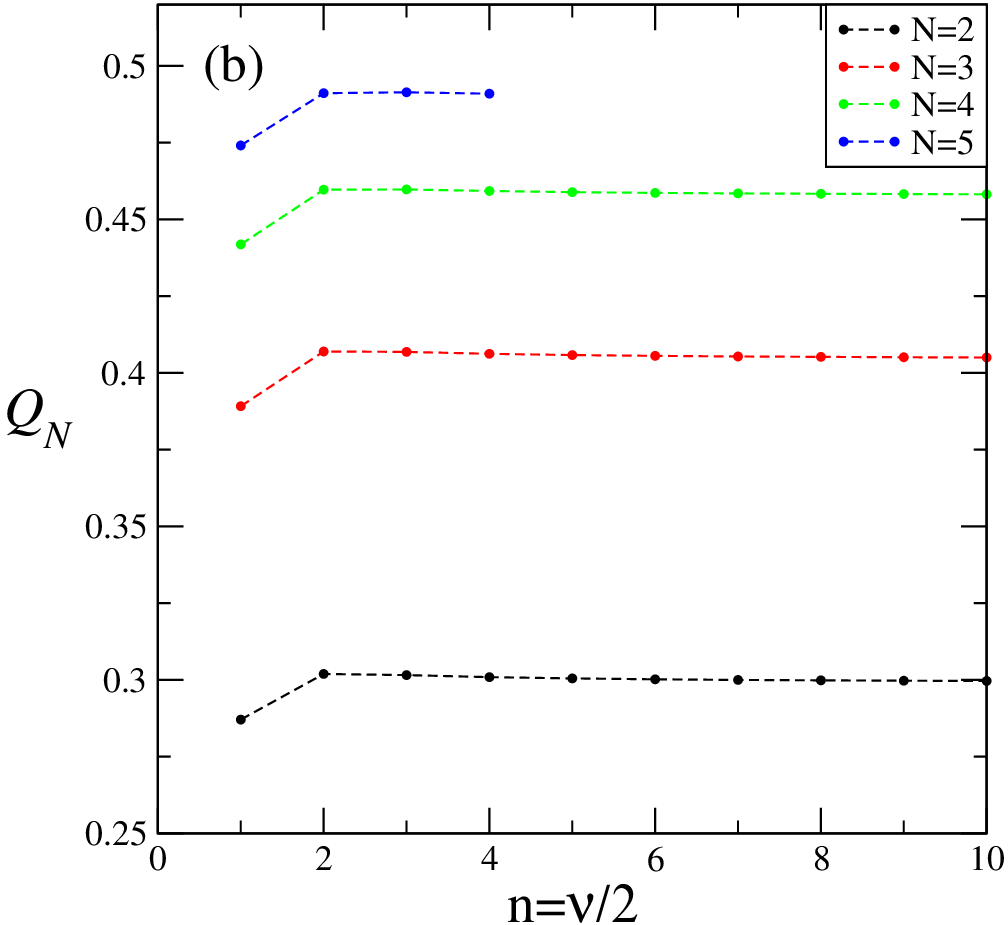}
\end{center}
\caption{  \label{fsvnN} a) von Neumann entropy  $S_N$, 
and b) JRW sub-entropy $Q_{N}$ for $N=2, 3, 4, 5$. Inset: The von Neumann 
entropy always shows a maximum for $\nu/2\in\left[2,4\right]$, to visualize it 
we plot a detailed view of the different data sets. The peaks can be shown 
together by subtracting a constant quantity to each data set.}
\end{figure}

The von Neumann entropy of the 1-RDM is a increasing function of the particle
number as can be appreciated in figure~\ref{fsvnN}a). All the curves shown in
figure~\ref{fsvnN}a) have a maximum, and  $S_N <
S_{N+1}<S_{N+2}<\ldots$ irrespective of the values of the interaction 
parameter $\nu$. The maximum of the von Neumann entropy seems to appear for 
$\nu/2\in\left[2,4\right]$ irrespective of the number of particles considered, 
but since we are showing only those values that can be obtained analytically, it 
is quite possible that if $\nu$ can be varied continuously that the actual 
maximum is reached for a non-integer value of $\nu$ that depends on the number 
of particles. 

From the data shown in figure~\ref{fsvnN}a), it can be appreciated that the
difference between successive values of the von Neumann entropy for a given
value of $\nu$ decreases when $N$ is
increased. Anyway, if the data is converging to some limit the increasing
difficulty to evaluate the elements of 1-RDM and its eigenvalues prevented us
from further exploring of this possibility.

As has been said in the Introduction, there is a number of entropy-like
functions that give information about the information content of the quantum
states under study. We choose to explore the Jozsa-Robb-Wootters
(JRW) sub-entropy \cite{Jozsa1994}, which is defined by

\begin{equation}
\label{eqjrw}
Q_{N}\,=\,-\sum_{k=1}^{(2 n (N-1)+1)^p}\,\left( \prod_{j\ne k }
\frac{\lambda_k}{\lambda_k-\lambda_j}\right)
\lambda_k\,\log_2(\lambda_k) \;.
\end{equation}
since it gives a rigorous lower bound for the accessible information contained
in the quantum state. This sub-entropy has been exactly calculated recently for
a variant of the two-particle Moshinsky Hamiltonian \cite{gn13}.

In figure \ref{fsvnN} b) we show $Q_{N}$  for the Calogero model for $N=2, 3,
 4, 5$. 
Note that  $Q_{N}$ qualitatively similar to $S_N$, but   $Q_{N}<<S_N$ in all the 
calculated values.

Interestingly, the JRW sub-entropy also shows the non-monotonous behavior shown
by the von Neumann entropy and  is too an increasing function of the particle
number. 

\section{The two-fermion Calogero model}

For two fermions we have an anti-symmetrical wave function

\begin{equation}
\label{ewf-f}
\Psi_0^{(F)}(x_1,x_2)\,=\,C_\nu\,sign(x_{1}-x_{2})\, 
\left|x_1-x_2\right|^\nu\,
e^{-(x_1^2+x_2^2)/2}\,,
\end{equation}

\noindent where $C_{\nu}$ is a normalization constant. 
 We note that, in this case we can ignore the absolute 
value when $\nu$ is a odd integer, and it is possible to calculate the 
normalization constant, then, for $\nu=2 n+1$ we have

\begin{equation}
\label{ewf-fodd}
\Psi_n^{(F)}(x_1,x_2)\,=\,2^n \sqrt{\frac{2 (2 n+1)!}{\pi (4 n+2)!}}\,
\left(x_1-x_2\right)^{2 n+1}\,
e^{-(x_1^2+x_2^2)/2}\,,
\end{equation}

Even when the wave function is antisymmetrical, the 1-RDM is a symmetric 
operator, then its matrix elements could be written  in a Hermite basis 
 following the steps of the bosonic case.

In this case the the 1-RDM matrix is $(2 n+2) \times (2 n+2)$, decomposable in 
an even and an odd  $(n+1) \times (n+1)$ blocks.
For the case $n=1 \Rightarrow \nu=3$ the matrices are  $2 \times 2$  and  its 
eigenvalues can be calculated analytically. 
The complete matrix is

\begin{equation}
\label{ern2f}
\left[\rho^{(1)}_{2}\right]\,=\,\left( \begin{array}{cccc}
 \frac{7}{20}&0&\frac{3}{10 \sqrt{2}}&0\\
 0&\frac{9}{20}&0& \frac{1}{10}\sqrt{\frac{3}{2}}\\
 \frac{3}{10 \sqrt{2}}&0&\frac{3}{20} &0\\
0& \frac{1}{10}\sqrt{\frac{3}{2}}&0& \frac{1}{20}
\end{array} \right) \,,
\end{equation}

\noindent whose eigenvalues are

\begin{equation}
\label{eavn2f}
\lambda_\pm\,=\,\frac{5\pm\sqrt{22}}{20}\,,
\end{equation}

\noindent both with multiplicity 2. This is a general property that holds for 
all $n$ because both, the even and odd $(n+1)\times(n+1)$ blocks of the 
1-RDM are different but isospectral matrices.

\begin{figure}
\begin{center}
\includegraphics[width=10.cm]{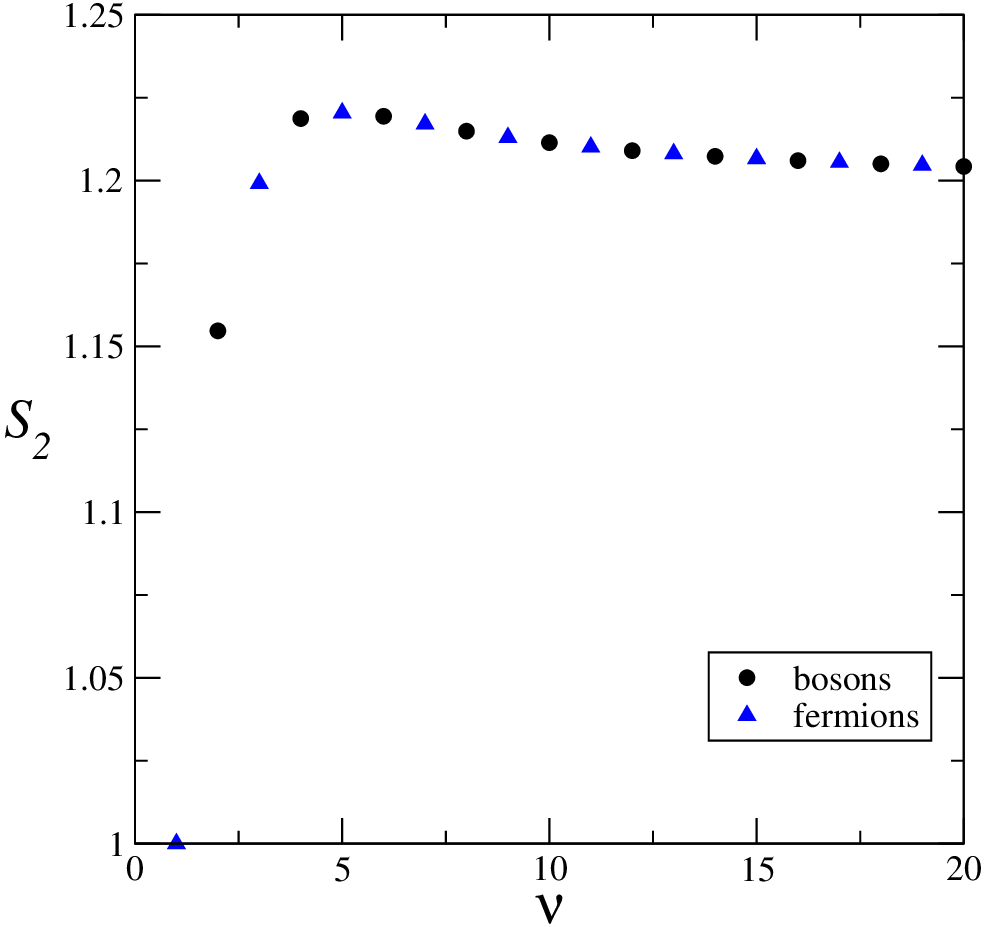}
\end{center}
\caption{  \label{fvnebf} von Neumann entropy for $N=2$ for bosons ($\nu=2 
n$, solid black dots) and fermions ($\nu=2 n+1$, solid blue triangles).}
\end{figure}

The  von Neumann entropy for two fermions is shown in figure \ref{fvnebf}, the 
bosonic entropy 
was also
included for comparison. Note that, at first and despite their closeness, both 
sets of points belong to different curves (the 1-RDMs are different) 
yet, because the strong repulsion between 
particles, the boson and fermion properties are almost coincident for large 
values of the coupling strength $\nu$. The quantitative differences 
between eigenvalues are largest in the neighborhood of the non-interacting 
limit $\nu \rightarrow 1$. This behavior is opposite to the one observed in the 
Moshinsky model where the difference between bosons and fermions is most 
noticeable for strong coupling~\cite{Schilling,Benavides}.

\section{Discussion and Conclusions}

As we have said previously, since the work of Moshinsky dealing with how much an
approximate two-particle wave function actually resembles the exact solution of
the problem, the need of  benchmarks were an approximation scheme to obtain the
information content of a problem can be tested has become more and more
pressing. In CV problems the usual criteria used to qualify the accuracy of a
given approximation are, basically, spectral, {\em i.e} if the approximate
eigenvalues found using the approximation are accurate in some sense then it is
assumed that the wave function and its information content should be accurate
too. We think that the exact results presented in this work can contribute as a
benchmark where to test some approximation schemes. 

Katsura and Hatsuda, some years ago, have obtained an exact 
formal expression for the $p$-RDM of the $N$-particle Calogero-Sutherland model 
on a ring of finite perimeter \cite{Katsura2007}. They were able to write the 
$p$-RDM as a sum of products of functions where, in each term, the dependencies 
with both sets of variables of the $p$-RDM is factorized, as in 
equation~(\ref{enrdm}). In that equation, the $p$-RDM depends on the 
sets of variables 
$\left\lbrace x_i\right\rbrace_p$ and $\left\lbrace y_j\right\rbrace_p$. 
Surprisingly, in the expression of Katsura and Hatsuda, each term contains the 
exact ground state of their model, $\psi_0^{CS}$, through products of the form 
$\overline{\psi_0^{CS}( \left\lbrace 
x_i\right\rbrace_p) } \;\psi_0^{CS}( \left\lbrace y_j\right\rbrace_p)$, instead 
of 
the completely factorized expression shown in equation~\ref{enrdm}, where each 
term depends on  a product of Hermite functions that depend in one, and only 
one, of the variables $x_i$ or $y_j$. This fact, owed to the Vandermonde 
determinant, allows us to find much more tractable expressions for the $p$-RDM 
than those found by Katsura and Hatsuda. If $N$ goes to infinity, both 
expressions, ours and the one of Katsura and Hatsuda, become formal since its 
evaluation becomes extremely troublesome. Although they provide an upper-bound
for the entropy in this case, it is valid only in the thermodynamic limit.

Despite that the present works deals with the one-dimensional Calogero model,
its extension to three dimensional problems with zero angular momentum is
rather direct. As a matter of fact, using the results presented above we
could obtain exactly some particular values of the von Neumann entropy for the
three dimensional two-particle Crandall atom. This model was studied by Manzano
{\em et al.} \cite{Manzano2010}, but in their work the von Neumann entropy was
calculated using an approximate Monte Carlo integration scheme. From our data we
observe that the work of Manzano {\em et al.} predicts  higher but very accurate
values for the von Neumann entropy in those cases where our result applies.

On the other hand, it is interesting to note that a non-monotonous behavior for
the von Neumann entropy was obtained for two-electron models using perturbation
theory when the interaction between the electrons is strongly localized and
weak \cite{Majtey2012}. 

The von Neumann entropy of the two-particle three dimensional case is a
non-decreasing function of the interaction strength, in contradistinction with
the one dimensional case. We think that this is so because the particle 
``impenetrability'', which is typical of one-dimensional problems, does not 
allow the particles to access some spatial regions, while
higher-dimensional problems do not posses this property. So far, we do not
have a proof confirming this hypothesis. Work around this lines
is in progress.

\acknowledgments

We acknowledge
SECYT-UNC and   CONICET for partial financial support. A.O.
warmly acknowledges the hospitality
of Universidad Nacional de C\'ordoba, where the work was done.



\begin{thebibliography}{20}
\bibitem{Chiribella2014}G. Chiribella and G. Adesso, Phys. Rev. Lett {\bf 112},
010501 (2014).
%
\bibitem{Braunstein2005}S. L. Braunstein and P. van Loock, Rev. Mod. Phys. 
{\bf 77}, 513 (2005).
%
\bibitem{Cerf2007}{\em Quantum Information with Continuous Variables of Atoms
and Light}, edited by N. Cerf, G. Leuchs, and E. S. Polzik
(Imperial College Press, London, 2007).
%
\bibitem{Tichy2011}M. C. Tichy, F. Mintert and A. Buchleitner, J. Phys. B: At.
Mol. Opt. Phys. {\bf 44}, 192001 (2011).
%
\bibitem{Moshinsky1968} M. Moshinsky, Am. J. Phys. {\bf 36}, 52 (1968).
%
\bibitem{Calogero1969} F. Calogero, J. Math. Phys. {\bf 10}, 2191 (1969).
%
\bibitem{Amovilli2004} C. Amovilli and N. H. March, Phys. Rev. A {\bf 69}, 054302
(2004).
%
\bibitem{Wootters1998}W.K. Wootters, Phys. Rev. Lett. {\bf 80}, 2245 (1998).
%
\bibitem{Amico2008}L. Amico, R. Fazio, A. Osterloh, and V. Vedral, Rev. Mod.
Phys. {\bf 80}, 517 (2008).
%
\bibitem{Vedral2002}V. Vedral, Rev. Mod. Phys. {\bf 74}, 197 (2002).
%
\bibitem{Nielsen2000}M.A. Nielsen and I. L. Chuang, {\em Quantum Computation and
Quantum Information} (Cambridge University Press, Cambridge, 2000).
%
\bibitem{Bennett1996} C.H. Bennett, G. Brassard, S. Popescu, B. Schumacher, J.
A. Smolin and W. K. Wootters, Phys. Rev. Lett. {\bf 76}, 722 (1996).
%
\bibitem{Horodecki1998} M. Horodecki, P. Horodecki and R. Horodecki, Phys. Rev.
Lett. {\bf 80}, 5238 (1998).
%
\bibitem{Horodecki2001}P. Horodecki and R. Horodecki, Quantum Information and
Computation {\bf 1}, 45 (2001).
%
\bibitem{Giedke2003}G. Giedke, M. M. Wolf, O. Kr\"uger, R. F. Werner, and J. I.
Cirac, Phys. Rev. Lett. {\bf 91}, 107901 (2003).
%
\bibitem{Jozsa1994} R. Jozsa, D. Robb and W.K. Wootters, Phys. Rev. A {\bf 49},
668 (1994).
%
\bibitem{Nichols2003} S. R. Nichols and W. K. Wootters, Quantum Information and
Computation {\bf 3}, 1 (2003).
%
\bibitem{Iemini2013}F. Iemini and R. O. Vianna, Phys. Rev. A {\bf 87}, 022327 
(2013)
%
\bibitem{Killoran2014}N. Killoran, M. Cramer and M.B. Plenio, Phys. Rev. Lett. 
{\bf 112}, 150501 (2014).
%
\bibitem{Gottlieb2005}A.D. Gottlieb and N.J. Mauser, Phys. Rev. Lett. {\bf 95},
123003 (2005).
%
\bibitem{Byczuk2012}K. Byczuk, J. Kune\v{s}, W. Hofstetter and D. Vollhardt,
Phys. Rev. Lett. {\bf 108}, 087004 (20012).
%
\bibitem{Zhang2014}J. M. Zhang and M. Kollar, Phys. Rev. A {\bf 89}, 012504
(2014).
%
\bibitem{Osenda2007} O. Osenda and P. Serra, Phys. Rev. A {\bf 75}, 042331
(2007). 
%
\bibitem{Ferron2009} A. Ferr\'on, O. Osenda, and P. Serra, Phys. Rev. A {\bf
79}, 032509 (2009).
%
\bibitem{Pont2010}F. M. Pont, O. Osenda, J. H. Toloza, and P.
Serra, Phys. Rev. A {\bf 81}, 042518 (2010). 
%
\bibitem{Majtey2012}A. P. Majtey, A. R. Plastino and J. S. Dehesa, J. Phys. A:
Math. Theor. {\bf 45}, 115309 (2012).
%
\bibitem{Li2008}H. Li and F. D. M. Haldane, Phys. Rev. Lett. {\bf 101}, 010504
(2008). 
%
\bibitem{Calabrese2008}P. Calabrese and A. Lefevre, Phys. Rev. A {\bf 78},
032329 (2008).
%
\bibitem{Manzano2010}D. Manzano, A. R. Plastino, J. S. Dehesa and T. Koga, 
J. Phys. A: Math. Theor. {\bf 43}, 275301 (2010).
%
\bibitem{Pruski72} S. Pruski, J. Ma\'ckowiak, O.Missuno, Rep. Math. Phys. {\bf 3}, 241 (1972).
%
\bibitem{ko13} P. Ko\'scik and A. Okopi\'nska, Few-body Syst. {\bf 54}, 1637
(2013).
%
\bibitem{Schilling} C. Schilling, Phys. Rev. A {\bf 88}, 042105 (2013). 
%
\bibitem{Benavides} C. L. Benavides-Riveros, I. V. Toranzo and J. S. Dehesa, 
Phys. B: At. Mol. Opt. Phys. {\bf 47}, 195503 (2014).
%
\bibitem{s71}  B. Sutherland, J. Math. Phys. {\bf 12}, 246 (1971).
%
\bibitem{fw08} P. J. Forrester and S. O. Warnaar, Bull. Am. Math. Soc. {\bf 45},
489 (2008).
%
\bibitem{lz10} S. K. Lando and A. K. Zvonkin, 
{\it  Graphs on Surfaces and Their Applications }, Springer (2004).

\bibitem{carlitz62} L Carlitz, Monatshefte f\"ur Mathematik {\bf 66},393 (1962).

\bibitem{mikhlin} see, for example, S.G. Mikhlin, {\em Integral Equations}, 
Pergamon Press (1957).
%
\bibitem{HA2015}  P. Ko\'scik, Phys. Lett. A {\bf 379}, 293 (2015).
%
\bibitem{HApc}  P. Ko\'scik, private comunication.
%
\bibitem{b91} S.D. Bajpai, http://www.sbc.org.pl/Content/33926/1992\_03.pdf.
%
\bibitem{gn13} M.L. Glasser and I. Nagy, Phys. Lett. A {\bf 377}, 2317 (2013).
%
\bibitem{Katsura2007}H.Katsura and Y. Hatsuda, J. Phys. A:
Math. Theor. {\bf 40}, 13931 (2007)
\end{thebibliography}
\end{document}